# Understanding Feasibility Study Approach for Packaged Software Implementation by SMEs

Issam Jebreen[*], Robert Wellington[1], and Stephen G. MacDonell[2]
SERL, School of Computing & Mathematical Sciences, Auckland University of Technology,
Private Bag 92006, Auckland 1142, New Zealand
*Corresponding author: issam.jebreen@aut.ac.nz, [1]rwelling@aut.ac.nz, [2]stephen.macdonell@aut.ac.nz

**Abstract**

*Software engineering often no longer involves building systems from scratch, but rather integrating functionality from existing software and components or implementing packaged software. Conventional software engineering comprises a set of influential approaches that are often considered good practice, including structured programming, and collecting a complete set of test cases. However, these approaches do not apply well for packaged software (PS) implementation; hence this phenomenon requires independent consideration. To explore PS implementation, we conducted ethnographic studies in packaged software development companies, in particular, to understand aspects of the feasibility study approach for PS implementation. From an analysis of these cases, we conclude that firstly; the analyst has more of a hybrid analyst-sales-marketing role than the analyst in traditional RE feasibility study. Secondly; the use of a live scenario software demonstration in order to convince the client to buy into the PS may lead to increased perceived feasibility and reduced resistance to PS implementation. Thirdly; the assessment criteria that are used to estimate the effort and time needed for PS implementation are new features, level of customization, software 'output', and technical needs. Fourthly; the feasibility study for PS implementation differs strongly from traditional RE as the analyst mainly considers how to deal with requests for modifications to existing functions.*

## 1. INTRODUCTION

Often, Small to Medium Sized Software Development Companies (SMSDCs) or small medium enterprises (SMEs) are unable to apply Requirements Engineering (RE) methods and techniques without modification [1]. In addition, shortcomings in applying RE methods due to time constraints or limited resources may arise [2]. Biirsner and Merten [3] noted that researchers need to intensify the investigation of RE practices in SMEs, otherwise SMEs will waste effort in searching for methodical orientation and dedicated tool support. Normally, the people responsible for requirements in SMEs are ambitious, but suffer from a scarcity of resources, and their time for undertaking research and trying different methods is very limited. They need quick methodical improvement of requirements elicitation, documentation, communication and traceability as well as more continuity of requirements management through the whole software lifecycle.

Karlsson [4] provides us with a summary of RE issues from studies into software development companies. However, none of these focus primarily on Packaged Software (PS) development and implementation. Furthermore, in most of these studies, the projects and organizations under consideration are large, in terms of the number of persons, the requirements involved, and the duration of the projects. Quispe [5] highlighted that there is a lack of knowledge about RE practices in SMEs. This lack of knowledge is particularly apparent when it comes to PS companies. It is difficult for researchers to gain much knowledge about how SMEs carry out RE given that SMEs seldom request external support. However, RE research should eventually enable those companies to become aware of state of the art or innovative RE techniques and to be able to improve their RE practice without external help [2]. Several questions remain unanswered. An important one being: How does requirements engineering in packaged software implementation contexts differ from traditional RE? In particular, there is a need for greater understanding of the feasibility study approach for PS implementation at SMEs.

The paper is organized as follows. In Sect. 3.2 we review literature related to our work. In Sect. 3.3 we briefly describe the research method. In Sect. 3.4 we present our findings and results, which are then discussed in Sect. 3.5. Finally, Sect. 3.6 sets out our conclusion and considers future work.

## 2. LITERATURE REVIEW

The concept of PS is defined as a ready-made software product that can be obtained from software companies, and which generally requires modification or customization for specific markets. They are often exemplified by enterprise resource planning (ERP) systems [6]. Previous researchers have highlighted that there is a lack of knowledge about the RE practices that assist PS implementation in these types of companies, and due to the particular characteristics of SMEs, several software engineering researchers have argued that most current RE practices are unsuitable for SMEs [5].

Table 1: Coding process

| Data extract | Codes for |
|---|---|
| Question: Tell about software demonstration? | Present software |
| The software we present is based on the notes from the sales team about the user's interest in potential software. Then we present the functions of the software that supports the user's business.... I think that helps the user know what their expectations could be for the software functions | Sales team report User's Interest Explain software functions Users' business Support Help users Users' expectations |
| It is good for us to make a software demonstration, in which we start to present a possible solution for users' issues. The flexibility that we want to have during software demonstration was constrained by a time limit since we only have one hour and a half to present our software....so we have to do our best to explain our software functions to the users. | Benefits of software demonstration Present a possible solution Users' Issues Constraints Time limitation Present software Explain software functions |

The poor use of RE practices (or the use of unsuitable practices) has often been identified as one of the major factors that can jeopardize the success of a software project [1, 7]. Whereas, it has also been recognized that following appropriate RE practices contributes to the success of software projects [8]. For example, Aranda [1] stated that gathering and managing requirements properly are key factors when it comes to the success of a software project. There is a general consensus that RE practices plays a very important role in the success or failure of software projects [2]. However, it is not possible to improve RE practices until areas that need improvement in an organization's current RE practice have been identified [3, 5].

The conclusion of the first Workshop on Requirements Engineering in Small Companies (2010) [3] was that existing RE techniques are not sufficient for small companies. However, size is not the only measure to categorize smaller companies and be the focus of research, that tacit knowledge and social structures in place in SMEs may play an important role in RE research, that introducing RE methods designed for larger companies may actually be harmful to the specific features of an SMSDC, and that RE methodologies need to be made more lightweight.

Much previous research on RE practices at SMEs has investigated the development of bespoke software. The majority of research in this area has related to soft- ware development studies [2, 7]. On the other hand, there are some studies that relate to the software development of packaged solutions, such as those by Daneva [9], Barney [10], Daneva and Wieringa [11]. However, these studies address PS development rather than the RE practices involved with PS implementation. Little attention has been paid to the phase of software package implementation from the perspective of the RE practices that are involved.

Further research addressing the topic is needed. Knowledge about this topic could be broadened and enhanced by carrying out intensive and in-depth work place studies.

This paper presents a study about SMEs, carried out in PS implementation and, more specifically, the RE practices perspective, highlighting some of the dynamics and complexity that these SMEs face, as well as their reactions to the challenges. Putting the organization and organizational practices at the centre of attention, this research advances our understanding of PS implementation from a work organization point of view, and in terms of RE practices.

## 3. RESEARCH APPROACH

An ethnographic study was conducted for 7 months across two software development companies. The business of the companies considered in this study is dominated by the provision of PS solutions. Data was collected throughout the research during field work. The three data collection methods, namely, interviews, participant observation, and focus groups, were used due to their suitability for qualitative research. Due to the hermeneutic characteristics of the analysis, the cases are not presented separately here, rather, every piece of data was analyzed in context, and the data relates not just to the context of the company, but to the specific individuals, projects, roles, process, and so on.

### 3.1 Data Analysis

Inductive analysis, used in this study, refers to an approach that primarily uses detailed reading of raw data to derive concepts, themes, and models through the researcher's interpretations of the raw data [12, 13]. During ethnographic research the ethnographer goes through a learning process at the same time as conducting their research. This process allows any prior presumptions that are found to be false to be redefined, reformed, or discarded [12]. We were then more open to experiencing what was going on around us, to paying attention to the details of the process, and to observe what was actually happening in the companies, rather than trying to search for relevant data. The first author conducted the field research and began to reflect on what actually occupied the analysts rather than on his own ideas or presumptions as a researcher. In field notes, he has noted information about everyday activities as well as conversations with the employees in the companies.

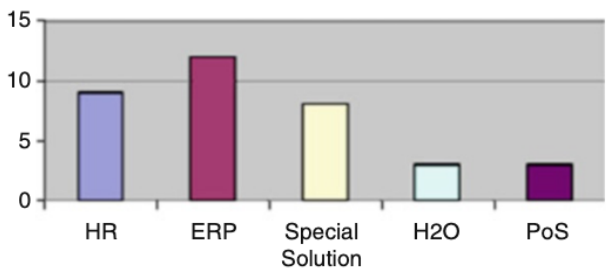

Figure 1: Number of project 15 cases observe

In this study, an initial round of field observations was conducted to find interesting topics involved with company's practices for PS implementation. We wished to discover the situations in which PS implementation occurs and understand the process that participants apply. After the initial field work, an initial round of coding was conducted in order to single out the descriptive and interpretive codes [13] (Table 1).

### 3.2 Participants
The two software development companies that participated in this research were established in 1997 and 1998. There were a total of 40 employees across the two companies, including people working in marketing and sales, analysts, developers, and management teams. The services they offer include software development, systems integration, and software localization.

There were a total of 16 research participants that included analysts and developers. The team leaders made up seven of the participants. Most of the participants were system analysts and developers at same time. The majority of the participants had a total experience of 3–10 years in the field, while a few had over 10 years' experience. Most of the participants had experience working as analysts, designers, and developers at the same time. Some participants had experience as analysts only, and some as developers only. Most participants had experience with business application software and database system software.

A total of 35 project cases were observed across the two companies. The PS types were Human Resource (HR) software, Enterprise Resource Planning (ERP) software, Special Solution software (such as a school management system), Restaurant Management software, and Point of Sale (PoS) software. Figure 1 below visually represents the number of cases observed.

## 4. FINDINGS AND RESULTS
This section is an ethnographic account of the events taking place when feasibility is studied for PS implementation. In doing so, we address a pre-implementation stage that includes (1) analysts' roles; (2) software demonstration utilizing a live scenario; and (3) mechanisms of scoping and creating packaged software offer. We also make a comparison between traditional RE and pre-implementation PS RE, given that the pre-implementation stage in this study resembles such feasibility studies as those used in traditional RE at a high abstract level. This is because feasibility studies in traditional RE and the pre-implementation stage of PS discussed here are similar in terms of their purpose, such as dealing with software objectives, time and budget.

### 4.1 Analysts' Roles
The analysts always carried out a software demonstration, wherein the analysts' roles have changed from just collecting requirements and undertaking software analysis to also engaging in the pre-sale of the software. It seems that this was likely due to the belief that analysts were the only ones who knew how the software was built, how it works, and how to explain it. It also appears that it was thought more likely that the client would buy the software if he/she feels the software solves their problems and helps them achieve their objectives — and the analysts may be better at explaining such issues than the sales team.

*"Before, the sales team made a software demonstration. But we found out that they just talked without understanding the software so sometimes they mentioned the wrong things to our client. The sales team is useless when making a software demonstration because they did not develop the software".* [General Manager]

Analysts approached the software demonstration from the perspective of convincing the client to buy the software, and from the perspective of identifying mis-alignments. This requires skill, knowledge, and experience in packaged software functionality and how components of the software relate to each other. It appears that, in one organization at least, the sales team did not possess such capabilities.

### 4.2 Software Demonstration Utilizing a Live Scenario
It can be observed that analysts consider the importance of a software demonstration from two dimensions: the business dimension which consists of presenting a possible solution to the client's issues and convincing the client to buy the software; the other dimension relates to software analysis, as the client might recognize new requirements and features in addition to those they initially perceived as requirements. The team leader of the analysts explained that the software demonstrations were usually capable of delivering the possible solution that the client needs, and the analysts believed that the system capabilities could be most effectively demonstrated with a live case.

*"We represent our software to a client by developing a real case scenario that can simulate and cover various aspects of a real situation within the client's work environment. It really helps us to explain our software functions and connect these functions to a real case".* [Team Leader]

One observed case of a live scenario was when a team of analysts demonstrated the H2O software 'Restaurant management system' for clients. The reason behind the live scenario was that H2O software included hardware functionality, such as using a personal digital assistant (PDA) to take customer orders, and software functionality, such as representing menu items.

During the demonstration process analysts explained what the software could do in order to solve the client's issues, such as their inability to follow an order, missing orders in the kitchen, and difficulties with the inventory of items. The analysts listed a set of functions that the software provided, such as making it easy to take an order, making sure the kitchen receives the order, helping the cashier to receive the payment required, and creating a record of the payment. The team of analysts then sought to represent this functionality by using a live scenario that allowed them to link software functionality to the business case to help convince the client about the software product. Meanwhile, initial requirements were also collected. In this case, the client had two different types of menu: one for local customers and another for tourists. The client also had three types of service: takeaway, delivery, and internet service. After the live scenario demonstration the client asked for each different kind of service to have its own printer in the kitchen, so that when the PDA was used it would not send all of the orders to the same printer.

It was clearly beneficial to plan the software demonstration by focusing on the client's specific business issues and to develop a live scenario, as this best showed that the analysts had a possible solution to the client's business process needs, and in this case, the client accepted the software offer after the demonstration.

### 4.3 Mechanisms of Scoping and Creating a Packaged Software Offer

The scoping process involves software analysis through discussion of high level modification requirements and new features. Without scoping, the software implementation time frame might expand to the degree where this would impact negatively on the cost and time involved in PS implementation. A participant explained:

*"It was a big issue, and it had required a lot of effort and time to work with that undefined scope to understand business practices and business requirements. Therefore, our strategy now is to define the scope of the software early on so that we will be prepared for the next step in the case that clients accept the software offer". [Team Leader]*

For example, a team of analysts demonstrated a Human Resource Management System (HRMS) to their clients. HRMS deals with demographic data, current employment information, employment history, qualification tracking, and salary information. The analysts' team and first author went to meet the client to discuss her needs. The client already had HR software and was experienced in its use. In this case, the analysts started by asking her about the issues that she and her employees faced with the current software. We discussed the client's issues and the possible solutions.

After the visit, the analysts' team provided an assessment report to the general manager. This report included a discussion of the client's issues, the modifications required to existing functions, and new features to be added. The client's issues were categorized into various types: transaction issues, such as employees' bonuses mechanism and costs related to the provision of uniforms, and output format issues, such as the software reports format. The general management and the analysts' team met to discuss possible ways of improving the already existing software in order to fit with the initial requirements, and to discuss the analysts' expectations of the development time-frame involved. After this, they also resolved client issues related to the price of the software (this is to make the software offer). Hence it is clear that one part of creating a software offer is scoping through software analysis. During this process of creating a PS offer, core requirements are emphasized but detailed requirements are neglected.

When creating a packaged software offer the analysts consider the scope of the offer according to the initial requirements, the number of modifications requested by the client, the nature and extent of the modifications, and the technical requirements involved. By focusing on such elements of a PS offer, the analysts were able to work towards their main goal of accurately estimating the cost and time requirements of PS implementation.

*"After we understand clients' need if it transaction function modifications that require 'customization' or new features, we will assess the cost and time require to develop such required". [Team Leader]*

When creating a PS offer, the software company also considers issues related to various kinds of 'assessment criteria' that were used to measure the level of impact on effort needed to develop, customize, and modify the PS. When we outline the relationship between the type of elements and the assessment level, this type of relation also highlights the software company's duties and the client's duties. The general manager and the analysts have defined the following assessment criteria for PS offers:

- New Features required, that consists of developing new functions that change the existing package
- Customization, which consists of modifying the existing functions to fill the gaps between the functions offered by a software package and the client's needs
- Software output, that consists of creating new reports or modifying existing reports or screens
- Technical needs, which consists of assessing the client's infrastructure requirements such as hardware and software.

### 4.4 Traditional RE vs. Pre-implementation

Table 2 characterizes the feasibility study in traditional RE [15] vs. pre- implementation PS. There are a number of differences in practice, and differences of purpose, between the elements of feasibility studies carried out in traditional RE and for pre-implementation PS RE. Traditional RE and pre-implementation PS RE share similarities as both can be seen as comprised of the same kinds of elements, and as, to some degree they sharing similar objectives and being influenced by simi- lar business concerns and technical concerns. For example, both processes can be considered in terms of the same dimensions, which involve goals, the business dimension, the software analysis dimension, domain knowledge, assessment criteria, critical decision, and scoping factors. However, within these dimensions, important differences appear.

Table 2: Traditional RE feasibility study vs. pre-implementation RE feasibility study

| Elements | Traditional RE: feasibility study | Pre-implementation PS: feasibility study |
|---|---|---|
| Goals | Are the overall objectives of the organization satisfied by the proposed system? Can the system be developed with the proposed budget and timeline? | What are the client issues? What is the possible solution? Is the possible solution within the scope of the software company's domain? What are the cost and time required for a possible solution? |
| Business dimension | Worthiness of proposed system. | Instilling confidence in the client, securing business, and creating a software offer. |
| Software analysis dimension | Information gathering to assist the assessment of proposed system. | Information gathering to identify client's issues, new requirements and new features needed (if any) to assess cost and time for proposed solution implementation. |
| Domain of knowledge | The development organization and the customer can cooperate to ensure that the domain is understood. | The development organization has to be an expert in the domain. |
| Assessment criteria | Objectives of the organization are satisfied by proposed system. System is developed with the proposed budget and timeline. | A new features level, customization level and output level. |
| Critical decision | Considers the worthiness of the proposed system, or regards changes, development decisions, seclude and budget. | The possible solution is within the software company's domain. |
| Scoping factors | Budget, timeline, technical and development issues. | Packaged software assessment level, elements, and limitation of work domain, client organization size, and client's issues. |

The analyst concerned with carrying out RE for PS implementation will be concerned with accessing different information and meeting different objectives than the analyst concerned with building custom-made software. For example, when building a bespoke system, traditional RE will focus on identifying whether the timeline and budget that have been proposed are feasible, and then with making sure that the organization's objectives can actually be met by the system that has been proposed. With pre-implementation of PS, however, the analyst must instead think about what the client's specific issues are and identify whether any existing pack- ages offered by the analysts' company can offer a solution. The analysts engaging in pre-implementation PS RE must also consider the possibility of refusing a request for a particular solution if that solution falls outside the scope of the company's capabilities or outside the scope of the company's current products. Part of the pro- cess of identifying whether the solution is within the company's scope may involve thinking about the time and cost involved with implementing a particular package or with making requested changes to that package.

With traditional RE, the main goal of the 'business dimension' of RE is concerned with establishing whether the proposed system is 'worthy': whether it can be created and whether it will actually satisfy the demands of the business and be the best possible system for the business. The analyst carrying out pre-implementation PS RE, however, will be engaged with different concerns, such as actually selling the proposed packaged system to the client by showing them how the package operates and how it could fulfill their requirements. The analyst carrying out pre-implementation PS RE must actively instill confidence in the client, secure his/her company's business, and create a software product offer.

The software analysis dimension in traditional RE and pre-implementation PS RE is quite similar. The analysts in both forms of RE carry out a range of activities that find out the client's issues that need solving and that help them to find initial requirements. They will later need to follow up on such requirements by checking in case new requirements are needed or new features need to be added to the pro- posed solution. If new features are required, they will again need to assess the cost and time involved with such requirements. However, there are some differences between the two forms of RE. In pre-implementation PS RE, analysts need to con- sider the modifications to existing functions that have been requested by clients. However, such considerations do not concern analysts practicing traditional RE.

The level of domain knowledge required for the analyst engaging in these different forms of RE also differs. With traditional RE, the analyst can gain sufficient knowledge of the client's domain by interacting with and listening to the client. The client is more active in advising the analyst what is needed in the system. With pre- implementation PS RE, however, the client will expect the development organization to already be an expert in the domain and to offer them the best possible solution or a range of viable solutions.

The assessment criteria used to develop and implement the software also differ between traditional RE and pre-implementation RE. With traditional RE, the feasibility of the system is seen to depend on whether the objectives of the organization will be satisfied by the proposed system. If it is considered that they will be, the system will then be developed in accordance with the proposed budget and timeline. Pre- implementation PS RE involves its own set of assessment criteria, as outlined above. With traditional RE, the main critical decision that needs to be made usually relates to confirming the worthiness of the proposed system. Other critical decisions, or factors, may relate to changes to the proposed system, or to budgetary factors or company developments. The analyst engaging in pre-implementation PS RE will make a critical decision when deciding whether the solution needed by the potential client is within the domain of the analyst's company.

The last element of comparison between feasibility studies in traditional RE and pre-implementation PS RE are scoping factors. Again, the scoping factors involved in the two different forms of RE are not the same. In traditional RE, scoping is guided mainly by the budget that has been set for the project, and by its timeline, and also by technical and development issues. Pre-implementation PS RE practice differs from this, as scoping for packaged software is influenced by a number of elements, including assessment criteria, the packaged software offer, and the limitation of the work domain, the client's organization size, and the client's issues.

## 5. DISCUSSION

This study describes activities that should help analysts conduct or manage a feasibility study for PS implementation in terms of RE practices by highlighting, in particular: (1) analysts' roles during pre-implementation; (2) software demonstration utilizing a live scenario; (3) mechanisms of scoping and creating a packaged software offer; and (4) traditional RE vs. pre-implementation PS RE.

**5.1 Analysts' Roles During Pre-implementation**
The main new finding of this study in terms of the role played by analysts in the feasibility study/pre-implementation stage, was that they rather than the sales team usually carried out the task of conducting a software demonstration for the client. In fact, in one of the cases observed, the sales team had previously carried out a software demonstration and had provided wrong information about the software to the client. Therefore, adopting the policy of analysts carrying out software demonstrations has been a company strategy to reduce the risk related to sales team members trying to sell features or accept to add new features to customers that had not actually been developed yet — an action that basically forced the company to include those features [16, 18].

One major difference, therefore, between the role of the analyst in pre- implementation PS RE and in traditional RE is that in pre-implementation RE the analyst has greater involvement in demonstration of the solution, being expected to conduct such software demonstrations. The analyst involved in pre-implementation PS RE is also expected to have some understanding of business concerns and how to engage in marketing. While the analyst in traditional RE is generally limited to developing the software and then stepping back and allowing sales and marketing teams to pitch the product to clients, the analyst doing pre-implementation PS RE has more of a hybrid analyst-sales-marketing role and is required to have the soft skills needed for a software demonstration presentation, such as presentation skills, communication skills, and sales skills. The analyst is no longer only concerned with software analysis but also with the business dimension of creating software [7, 19].

**5.2 Software Demonstration Utilizing a Live Scenario**
One of the factors differentiating pre-implementation PS RE practices from traditional RE practices is that analysts have a choice of how to conduct a feasibility study. Analysts may be able to offer more than one solution to the client. In such a case, analysts then need to choose which solution is the preferred one to offer to the client [7]. In order to make such a decision, analysts hold meetings that involve themselves and the sales and marketing teams. The limitations of the work domain of the analysts' company are considered. Other factors that are considered relevant to making a decision about the solution to offer include the size of the client's organization, the number of users at the organization, the kinds of departments the organization has, and the kinds of transactions the client organization will need to carry out.

The live scenario was used in cases where it was decided that this provided the best option for showing the capabilities of a PS. The live scenario aims to simulate a situation that could occur in the client company's real work environment. It was found, therefore, that contrary to simply describing or demonstrating a software package during a meeting, the live scenario involved analysts creating an environment that simulates the client company's site in order to show the client a real-time and live example of how the system could function within their operational context.

It appeared that using a live scenario helped the analysts to better understand and respond to the needs of the client company. For example, by conducting the live scenario the analysts were better able to see the challenges actually faced by the company, it was found that this kind of software demonstration could have a strong influence on whether the client would purchase the solution [7, 18].

It was also found that the planning of such a live scenario software demonstration relied on the analysts developing soft skills such as being able to present in a way that is personable and convincing, not merely to display knowledge about software.

The use of the live scenario for pre-implementation PS RE is different from any procedure used by analysts in traditional RE because the analyst conducting traditional RE is not typically concerned with showing the client company how a soft- ware solution may work for them through the use of imagined scenarios or by analogies showing how the solution worked for another company. In traditional RE the analyst is only concerned with building a solution to meet the demands of the client and to test them, during the software's development, that it does meet their requirements or fix their problems, and adjust the software until it is correct. The use of the live scenario during software demonstration, therefore, is a method unique to pre-implementation PS RE.

### 5.3 Mechanisms of Scoping and Creating a Packaged Software Offer

The analysts attempted to define the scope of the software during discussions with potential clients about their needs. Analysts believed it was important to carry out this scoping process early on since this would help them to construct a software offer, since such scoping would help everyone involved to maintain control of the time taken for implementation. Collecting such information not only provided analysts with details about what the new software needed to do, but also helped them to see what its limitations would be and what new features or modifications would be necessary. This step therefore helped them significantly with designing a PS that would suit the client.

It was also observed that the steps the software company took related to software scoping were generally limited to finding out information about only the core requirements of the system or solution to be implemented. This was found to involve transaction issues and output format issues [17]; during the scoping process the analysts were not concerned with discovering detailed requirements.

### 5.4 Traditional RE vs. Pre-implementation PS RE

The scoping factors involved when creating bespoke software and therefore con- ducting traditional RE are; budget constraints, timeline issues and constraints, technical issues, and development issues. Analysts conducting traditional RE will consider whether the timeline and budget that have been set are feasible, and must also ensure that the client organization's objectives can be met by the proposed software. Their main concern is whether the system that is developed will be 'worthy' for use.

Analysts engaging in pre-implementation RE must think about the client's specific issues and decide whether any existing packages offered by their company can offer a solution. They will need to consider the time and cost involved with implementing a particular package and with making requested changes to that package, and they may well decide to refuse a request for a particular solution if that solution falls outside the scope of the company or outside the scope of their current products. In this regard, pre-implementation PS RE differs strongly from traditional RE as analysts practicing traditional RE do not need to consider how to deal with requests for modifications to existing functions. Neither does the analyst practicing traditional RE need to engage in scoping with the aim of creating a software offer or actively take part in selling the proposed packaged system to the client.

## 6. CONCLUSION

This ethnographic account of pre-implementation PS RE, for the first time, shows how a software development company of this size (small-medium) approaches the challenge of managing the pre-implementation process at this point of the packaged software implementation life cycle. We have highlighted elements and assessment criteria involved with creating a software offer that is based on modification and customization of existing packaged software.

In this paper, we have outlined the importance of the software demonstration, and as such that the role of the analyst engaged in pre-implementation PS RE differs from their respective role in traditional RE. In pre-implementation PS RE, the analyst is likely to be the staff member delivering the software demonstration. Because analysts know how the software is built, how it works, and how to explain it, clients may be more likely to buy the software when the analyst explains it to them. Clients will buy the software if they feel that the software will help them to achieve their objectives and to solve their problems. Therefore, the analysts' presentation of the software is approached from the perspective of convincing the client of the pack- age's suitability, and from the perspective of identifying misalignments, within the client's context. This presentation requires knowledge related to packaged functions and how functions are related to each other, as well as communication skills and the ability to persuade the client.

Future work should be undertaken to discover the differences in philosophy behind release plans for packaged software between large packaged software development companies and companies that are SMEs. From observations and from previous literature, it appears that large packaged software development companies tend to have very detailed release plans and schedules for future packaged software products mapped out months or years in advance, while SMEs may take a more ad hoc approach to release planning that instead involves continuous improvement of their product in response to clients' requirements and clients' responses to their product.

## REFERENCES

1. Aranda J, Easterbrook S, Wilson G (2007) Requirements in the wild: How small companies do it. In Requirements Engineering Conference, 2007. RE'07. 15th IEEE International (pp. 39–48). IEEE.

2. Merten T, Lauenroth K, Biirsner S (2011) Towards a new understanding of small and medium sized enterprises in requirements engineering research. In: Requirements engineering: founda- tion for software quality. Springer, Berlin, pp 60–65
3. Bürsner S, Merten T (2010) RESC 2010: 1st Workshop on Requirements Engineering in Small Companies. econstor, 128.
4. Karlsson L et al (2007) Requirements engineering challenges in market-driven software devel- opment—an interview study with practitioners. Inform Software Technol 49(6):588–604
5. Quispe A, Marques M, Silvestre L, Ochoa SF, Robbes R (2010) Requirements engineering practices in very small software enterprises: A diagnostic study. In Chilean Computer Science Society (SCCC) (pp. 81–87). IEEE.
6. Xu L, Brinkkemper S (2007) Concepts of product software. Eur J Inform Syst 16:531–541
7. El Emam K, Madhavji NH (1995) A field study of requirements engineering practices in information systems development. In: Proceedings of the second IEEE international symposium on Requirements Engineering, 1995. IEEE
8. Solemon B, Sahibuddin S, Ghani A (2009) Requirements engineering problems and practices in software companies: An industrial survey. In Advances in Software Engineering (pp. 70–77). Springer Berlin Heidelberg.
9. Daneva M (2004) ERP requirements engineering practice: lessons learned. IEEE SOFTWARE
10. Barney S, Aurum A, Wohlin C (2008) A product management challenge: Creating software product value through requirements selection. J Syst Architect, 54(6), 576–593.
11. Daneva M, Wieringa RJ (2006) A requirements engineering framework for cross-organizational ERP systems. Reqs Eng 11:194–204
12. Hammersley M, Atkinson P (2007) Ethnography: principles in practice. Routledge, London
13. Miles MB, Huberman AM (1994) Qualitative data analysis: an expanded sourcebook. Sage, Thousand Oaks, CA
14. Khoo HM, Robey D (2007) Deciding to upgrade packaged software: a comparative case study of motives, contingencies and dependencies. Eur J Inform Syst 16:555–567
15. Sommerville I (2004) Software engineering, 7th edn. Addison-Wesley, Boston, MA
16. Jantunen S (2010) The benefit of being small: exploring market-driven requirements engineering practices in five organizations. In: Proceedings of the 1st workshop on RE in Small Companies (RESC), Essen, Germany
17. Dittrich Y, Vaucouleur S, Giff S (2009) ERP customization as software engineering: knowledge sharing and cooperation. IEEE Software 26(6):41–47
18. Jebreen I, Wellington R (2013) Understanding requirements engineering practices for pack-aged software implementation. In: 4th IEEE international conference on software engineering and service sciences, Beijing, China
19. Soh C, Kien SS, Tay-Yap J (2000) Enterprise resource planning: cultural fits and misfits: is ERP a universal solution? Comm ACM 43(4):47–51